\newif\ifMNRAS
\MNRAStrue 

\PassOptionsToPackage{pdfpagelabels=false}{hyperref} 

\ifMNRAS
    \documentclass[fleqn,usenatbib,useAMS]{mnras}
    \usepackage{graphicx}
\else
    \documentclass[twocolumn]{aastex63}
\fi

\usepackage{acronym}
\usepackage{hyperref}
\usepackage{amsmath,amsfonts,amssymb}
\usepackage{mathrsfs}
\usepackage{mathtools}
\usepackage{multicol}
\usepackage[utf8]{inputenc}

\usepackage[normalem]{ulem}
\usepackage[dvipsnames]{xcolor}
\usepackage{xspace}

\usepackage{pifont}

\ifMNRAS
\title[Neutrino Signatures]{Neutrino Signatures of One Hundred 2D Axisymmetric Core-Collapse Supernova Simulations}

\author[Vartanyan et al.]{
 \href{https://orcid.org/0000-0003-1938-9282}{David Vartanyan$^{1}$}\thanks{E-mail: dvartanyan@carnegiescience.edu},
\href{https://orcid.org/0000-0002-3099-5024}{Adam Burrows$^{2}$}
\\
$^{1}$Carnegie Observatories, 813 Santa Barbara St, Pasadena, CA 91101, USA; NASA Hubble Fellow\\
$^{2}$Department of Astrophysical Sciences, 4 Ivy Lane, Princeton University, Princeton, NJ 08544, USA
}

\date{Accepted XXX. Received YYY; in original form ZZZ}

\pubyear{2023}

\else
\shorttitle{2D}
\shortauthors{Vartanyan et al.}
\fi

\begin{document}
\label{firstpage}
\pagerange{\pageref{firstpage}--\pageref{lastpage}}
\maketitle
\ifMNRAS
\else
\title{}

\fi

\begin{abstract}
We present in this paper a public data release of an unprecedentedly-large set of core-collapse supernova (CCSN) neutrino emission models, comprising one hundred detailed 2D-axisymmetric radiation-hydrodynamic simulations evolved out to as late as $\sim$5 seconds post-bounce and spanning a extensive range of massive-star progenitors. The motivation for this paper is to provide a physically and numerically uniform benchmark dataset to the broader neutrino detection community to help it characterize and optimize subsurface facilities for what is likely to be a once-in-a-lifetime galactic supernova burst event. With this release we hope to 1) help the international experiment and modeling communities more efficiently optimize the retrieval of physical information about the next galactic core-collapse supernova, 2) facilitate the better understanding of core-collapse theory
and modeling among interested experimentalists, and 3) help further integrate the broader supernova neutrino community. 
\end{abstract} 

\ifMNRAS
    \begin{keywords}
    stars - supernovae - general
    \end{keywords}
\else
    \keywords{
    stars - supernovae - general }
\fi

\section{Introduction}
\label{sec:int}

Core-collapse supernova theory has evolved significantly over the last six decades (\citealt{burrows2013, janka2012, muller2017, ott2018_rel, oconnor_couch2018b, vartanyan2018a, burrows_2019, burrows_2020, glas2019, 2021Natur.589...29B, vartanyan2020,vartanyan2022,tsang2022,wang}) and has reached a pivotal juncture in its history.  Previously, delayed by the complexity and computational resources necessary to address the relevant physics in sufficient and credible multi-dimensional detail, most sophisticated core-collapse supernova simulations now witness explosions without artifice. The turbulent neutrino-driven mechanism, at least in broad outline, is now the consensus mechanism for the vast majority of core-collapse supernova explosions \citep{janka2012,burrows2013,burrows_2020,2021Natur.589...29B}. This is not to say that there is no need for significant quantitative improvement. Moreover, there remain issues concerning various neutrino-matter interactions, the role of neutrino oscillations, and the actual (and determinative) structures of the massive-star progenitors. Nevertheless, from a slight conceptual distance, the theory, though complex, is looking realistic and can boast many detailed predictions.

Given these developments, we present in this paper a public data release of an unprecedentedly-large set of core-collapse supernova (CCSN) neutrino emission models, comprising one hundred detailed 2D-axisymmetric radiation-hydrodynamic simulations evolved out to as late as $\sim$5 seconds post-bounce and spanning a extensive range of massive-star progenitors\footnote{The full dataset can be found at: \url{http://www.astro.princeton.edu/~burrows/nu-emissions.2d.large/} and \url{https://dvartany.github.io/data/}}. Though some 1D simulation suites exist \citep{2012ApJ...757...69U,swbj16,2018MNRAS.475.1363H,suwa2019,2021ApJ...909..169K}, spherical-symmetry is insufficient to capture the convective nature of the CCSN that significantly affects the emergent neutrinos (\citealt{2006ApJ...645..534D, radice2017b}. Moreover, these simulations often use explosion prescriptions rather than sophisticated neutrino transport, or reduced the dimensionality to 1D to continue to several seconds post-bounce (e.g. \citealt{2021ApJ...915...28B}). 


A uniform suite of neutrino simulation data is essential to ensure that the neutrino experimental and theoretical communities are adequately equipped to optimize the retrieval of physical information about the next galactic core-collapse supernova. Such an event is a major secondary motivation for the flotilla of subsurface facilities emerging (e.g., Super-Kamiokande, SK, \cite{abe2016} or Hyper-Kamiokande, HK, \cite{abe2018}; the Deep Underground Neutrino Experiment, DUNE, \cite{ankowski2016,migenda2018}; the Jiangmen Underground Neutrino Observatory, JUNO, \cite{juno}; THEIA \cite{theia};
  and IceCube, \cite{abbasi2011,kopke2011}). The motivation for this paper is to provide a physically and numerically uniform benchmark dataset to this broader neutrino detection community to help it characterize and optimize subsurface facilities for this once-in-a-lifetime galactic epiphany. 

There exists neutrino-analysis software coupling CCSN simulations with neutrino signals, such as SNEWPY \citep{snewpy} and SNOwGLoBES \citep{snowglobes,nagakura2021a} within a collaborative framework (e.g., \citealt{snowmass,snews}). Moreover, early and joint detection of neutrinos with gravitational waves would provide information on the global dynamics of CCSNe, the turbulent radiation-hydrodynamic instabilities, and the compact remnant properties, providing constraints on the nuclear equation of state \citep{vartanyan2019}. 


In Sec.\,\ref{sec:methods}, we summarize the methods of this paper, including the models studied and a few simulation properties. In Sec.\,\ref{sec:results}, we discuss various example CCSN neutrino simulation diagnostics and correlations. One purpose of this data release is to enable others to find their own diagnostics, signatures, correlations, and systematics, but we kickstart such an effort with the modest set of examples. In Sec.\,\ref{sec:2D-3D}, we compare and calibrate results from analogous 2D and 3D simulations 
In Sec.\,\ref{sec:conc} we summarize the import of this reference neutrino dataset.

\section{Methods}\label{sec:methods}

To ascertain their explodability, we originally selected a subset of one hundred initial massive-star progenitor models at core-collapse from 9 to 26.99 M$_{\odot}$ and evolved them in 2D (axisymmetry) using our state-of-the-art code F{\sc{ornax}}. The progenitor models between 9 to 11.75 M$_{\odot}$ were selected from \cite{swbj16} and the models between 12 and 26.99 M$_{\odot}$ were selected from \cite{sukhbold2018}.
F{\sc{ornax}} \citep{skinner2019} is a multi-dimensional, multi-group radiation-hydrodynamic code constructed to study core-collapse supernova explosions and their byproducts. It features an M1 solver \citep{2011JQSRT.112.1323V} for neutrino transport with detailed neutrino microphysics and approximate general relativity \citep{marek2006}. We employed for this study 12 neutrino energy groups spaced logarithmically between 1 and 300 MeV for the electron neutrinos and to 100 MeV for the anti-electron- and ``$\nu_\mu$"-neutrinos (bundled as $\mu$ and $\tau$ neutrinos and anti-neutrinos) and the SFHo nuclear equation of state \cite{2013ApJ...774...17S}, broadly consistent with extant theoretical and experimental constraints on the nuclear equation of state \citep{2023Parti...6...30L,2021PhRvL.126q2503R,2017ApJ...848..105T}. 
The progenitor models and simulations are initially non-rotating, though some degree of rotation is naturally induced due to fallback \citep{2007Natur.445...58B,coleman}. 

The models have a resolution of 1024$\times$256 in $r$, $\theta$ with outer radii extending between 30,000 km for the lower mass stellar progenitors out to 100,000 km for the most massive progenitors.  These models were chosen to be representative as much as possible of the Salpeter initial mass function. They were selected to span broadly the distribution in density profiles, compositional interfaces, compactness, and Ertl parameter $\mu_4$/$M_4$ \citep{2016ApJ...818..124E}. Generally, we truncated the black hole forming models at $\sim$1 second post-bounce, when it was clear they had not exploded, and continued the exploding models out to $\sim$3$-$5 seconds post-bounce. We dump and provide the data every millisecond (ms) and angle-average the luminosities. At the dataset URLs (see footnote and Data Availability statement), we provide the luminosities, average neutrino energies, and RMS neutrino energies, as well as the full neutrino spectra, for all three species categories.

Of the 100 models, we find that 63 explode and 37 fail to do so, ultimately resulting in black hole formation. All models that fail to do so lie between 12$-$15.38 M$_{\odot}$, and within this mass range we find $\sim$70\% of models do not explode, all else being equal. Interestingly, CCSN simulations do consistently find models in this range are less explodable (\citealt{burrows_2019, burrows_2020,2021Natur.589...29B, summa2016,oconnor_couch2018a}).

The models presented comprise both the largest set of 2D-axisymmetric radiation-hydrodynamic CCSNe simulations published to-date and are out to the latest times for any such collection. This set is a precursor to an upcoming release of an analogous set of 3D CCSNe simulations out to very late times. The latter, however, requires substantially more computer time that perforce will delay its release. Concerning the general explosion quantities (neutrino energies and luminosities, explosion timing, and outcome), we expect roughly similar behavior in 2D and 3D. Of course, for explosion morphologies, energies, nucleosynthesis, and gravitational emission wave, one requires 3D simulations. However, for general comparative and diagnostic purposes, we suggest that 2D simulations are a robust compromise between explosion breadth (we can do many simulations out to later times in a relatively short period of time) and simulation fidelity.
The 2D models highlighted here (though not these neutrino data) have been published in other contexts in two earlier papers \citep{wang,tsang2022}; these provided both analytical/semi-analytical and machine-learning predictions of explosion outcome.


\section{Results: Example Trends and Correlations}\label{sec:results}
The main unifying principle guiding this release is that compactness \citep{2011ApJ...730...70O} provides a natural ordering of many physical quantities. As we will show, these include the neutrino luminosities, mean neutrino energies, and infall accretion rates. Compactness provides a stronger monotonic prediction of these properties than does the progenitor mass. These ideas are not new. However, we illustrate them here using the largest set ever assembled of detailed, two-dimensional, radiation-hydrodynamic CCSN models. In Table\,\ref{sn_tab}, we summarize the 100 models, indicate which explode, and show total runtimes, the compactness (at 1.75 M$_{\odot}$), and the explosion times.

We emphasize that these correlations do not indicate compactness is a good metric of explosion outcome, as has often been assumed (e.g., \citealt{sukhbold2018}). In fact, compactness is not a monotonic indicator of explosion at all $-$ models with low-compactness may not explode by the neutrino mechanism, while models with high compactness may explode \citep{burrows_2019,burrows_2020,wang}. Curiously, we find here and previously in print \citep{vartanyan2018a,burrows_2020,tsang2022,wang} that regions of intermediate compactness, which correspond to models between 12 and 15 M$_{\odot}$, are less likely to explode by the neutrino mechanism \citep{burrows_2019,burrows_2020}. 

In an earlier study \citep{tsang2022}, we found that a random-forest machine-learning approach was able to identify regions of intermediate compactness less likely to explode with $\sim$80$\%$ accuracy, and that this corresponded with analytical and semi-analytical estimates of explosion outcome \citep{wang}. We note that using the Silicon-Oxygen (Si/O) interface as an explosion metric performed better at predicting explosion outcome, with a $\sim$90$\%$ accuracy (see also \citealt{fryer1999,vartanyan2018a,summa2016,oconnor_couch2018b}).


\subsection{2D to 3D Comparison}\label{sec:2D-3D}

\begin{figure*}
    \centering
    \includegraphics[width=0.74\textwidth]{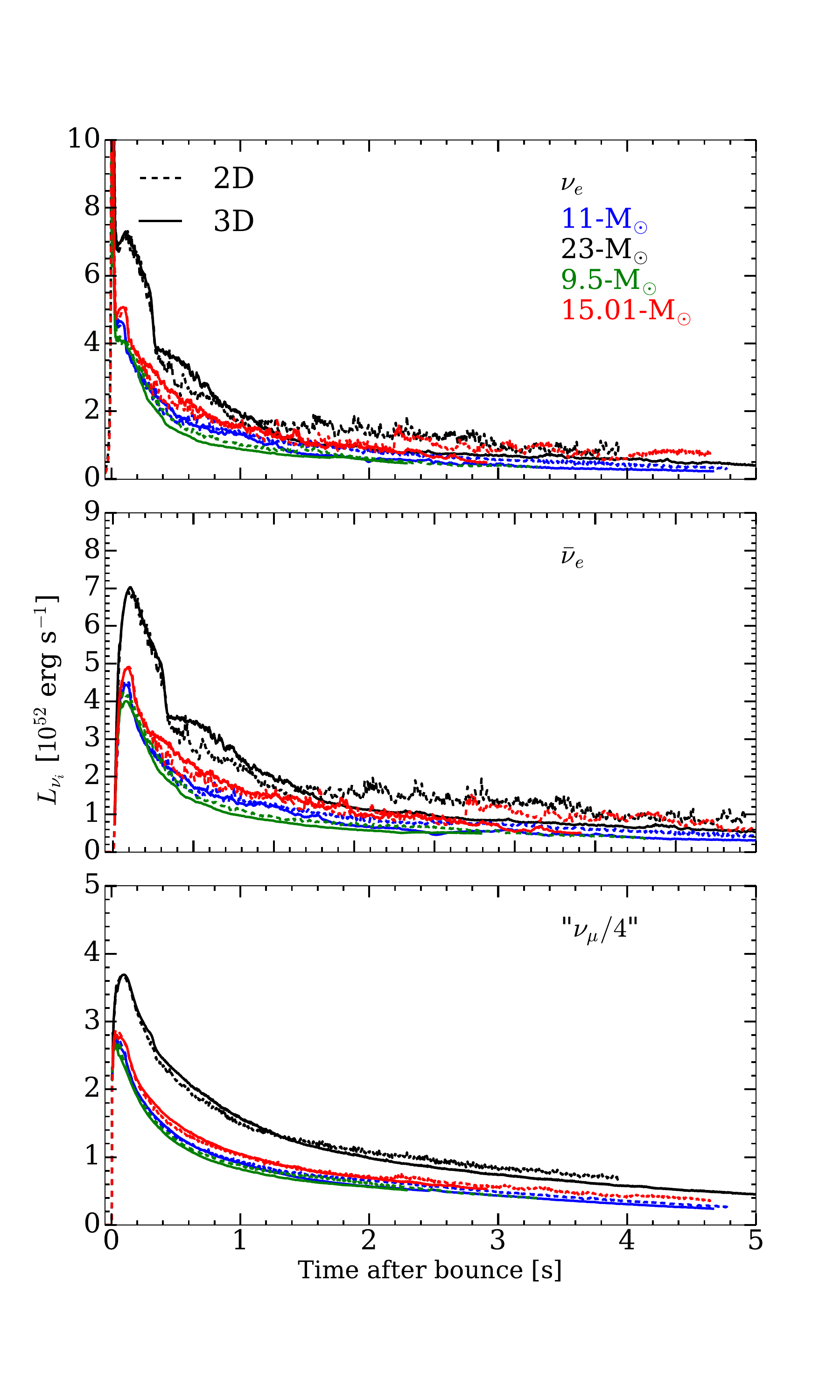}
    \caption{2D and 3D comparison of the neutrino luminosity at 500 km in the observer frame for all three species for exploding models 11- and 23-M$_{\odot}$.}
    \label{fig:lum-3D}
\end{figure*}

In Fig.\,\ref{fig:lum-3D}, we plot for a series of 2D models the neutrino luminosities and average energies, respectively, with their 3D counterparts. We emphasize that, generally, there is less short-timescale variability in angle-averaged 3D neutrino quantities than in the corresponding angle-averaged 2D neutrino quantities. We associate this with the artificial axisymmetric constraint in 2D. Note also that, in 3D, we expect typical angular variations by viewing angle of $\sim$15$\%$ in the neutrino quantities \citep{vartanyan2020}. At later times (beyond 1$-$2 seconds post-bounce), we see deviations in the electron neutrino and anti-neutrino luminosities for the more massive progenitors, with 2D simulations yielding higher luminosities as a result of the sustained higher accretion in 2D.

\subsection{A Few Example Compactness Correlations}

The compactness parameter characterizes the core structure and is defined as \citep{2011ApJ...730...70O}:
\begin{equation}
\xi_M= \frac{M/M_{\odot}}{R(M)/1000\, \mathrm{km}}\,,
\end{equation}
where the subscript $M$ denotes the interior mass coordinate at which the compactness parameter is evaluated. For our purposes, we evaluate the compactness parameter $\xi_{1.75}$ at $M$ = 1.75 M$_{\odot}$, generally encompassing the Si/O interface for many of the progenitor models. The compactness is often used as an ab-initio explosion condition because it depends only on the progenitor properties. While higher compactness is correlated with higher luminosities, accretion rates, and remnant masses \citep{oconnor2013},  compactness does not readily lend itself as an explosion condition, and suggestions that explosion is inhibited above a certain compactness parameter are likely false \citep{burrows_2020,wang}.  In all diagnostics indicated below, we provide ordering by both compactness and by progenitor ZAMS mass.

\begin{figure*}
    \centering
    \includegraphics[width=0.47\textwidth]{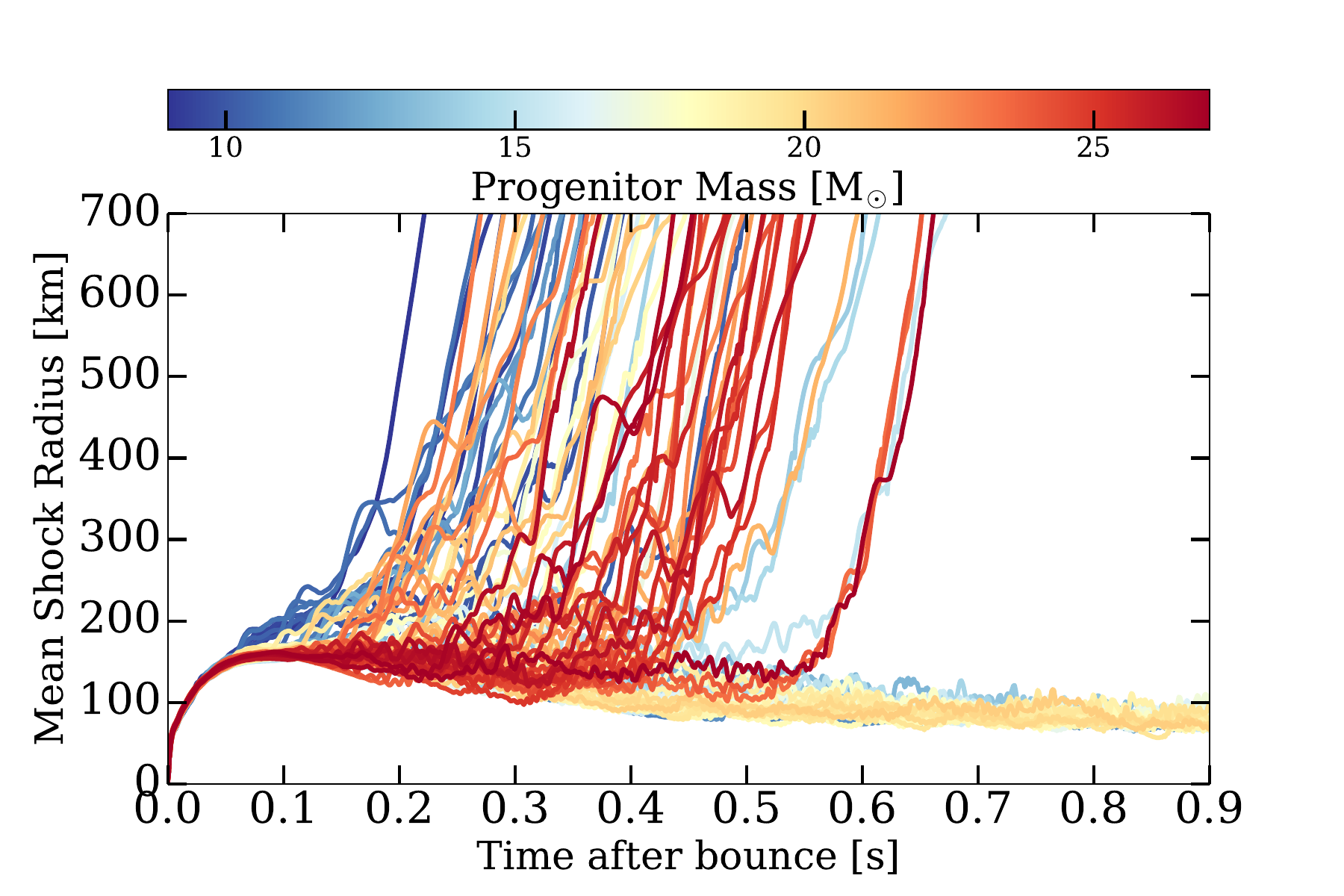}
    \includegraphics[width=0.47\textwidth]{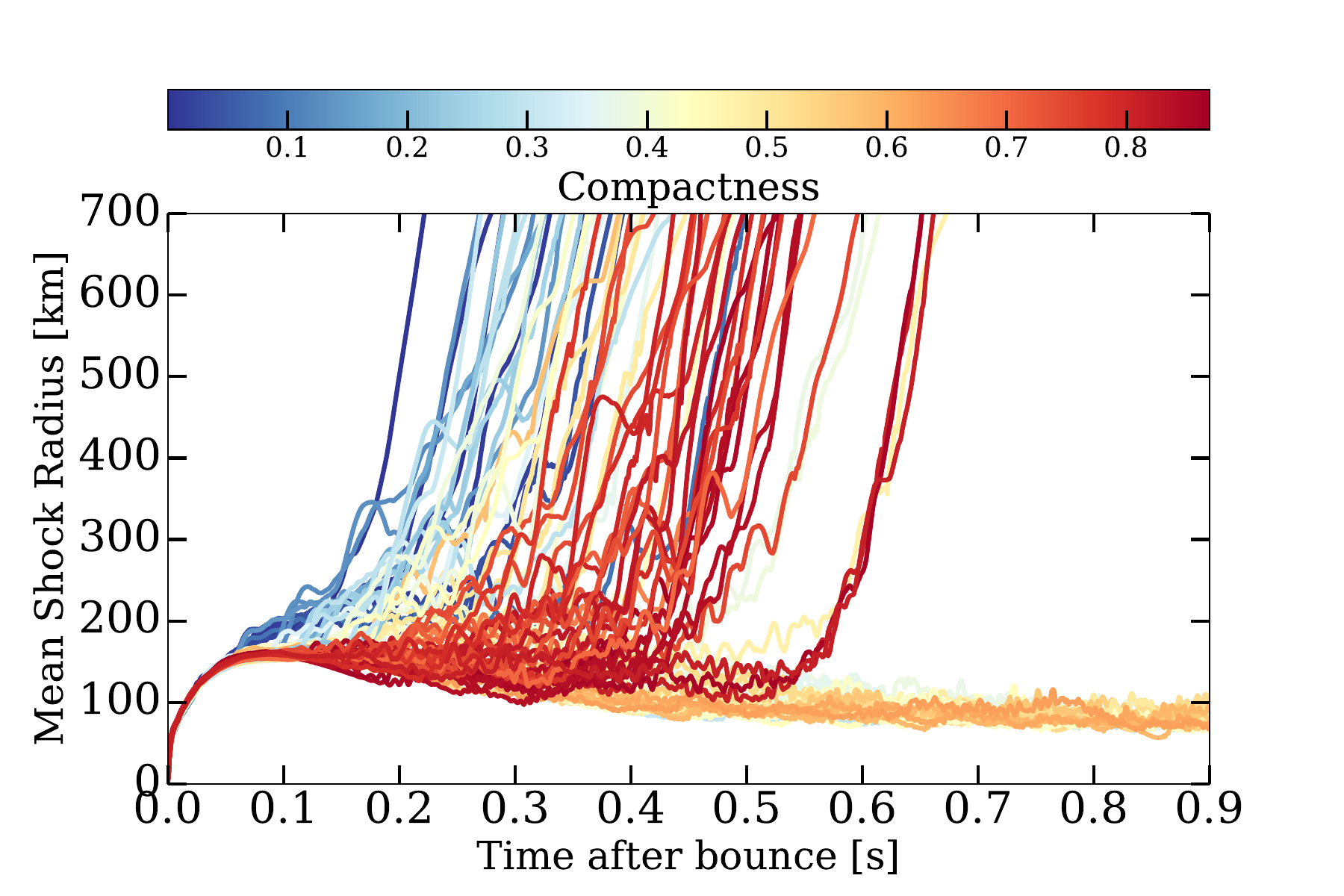}
    \caption{Mean shock radii (in km) as a function of time after bounce (in seconds) for the 100 models explored here as a function of both progenitor mass (\textbf{left}) and compactness (\textbf{right}). Note that, when ordering by compactness, we see generally that low- and high-compactness models explode, whereas some intermediate ones fail to do so (see also \protect\citealt{wang, tsang2022}).}
    \label{fig:rs}
\end{figure*}

In Fig.\,\ref{fig:rs}, we plot the mean shock radii as a function of time after bounce and as a function of both progenitor mass and compactness. Note that all models stall at $\sim$60 ms and at similar shock radii, with lower mass models reviving into explosion earlier. For models that do explode, we note a distribution of explosion time that is nearly monotonic with compactness and ranges from $\sim$120 ms post-bounce to $\sim$600 ms post-bounce. This range in explosion times corresponds to the timing of Si/O interface accretion $-$ for instance, model 26.99 M$_{\odot}$ has an Si/O interface near 2 M$_{\odot}$ at core-collapse, and hence accretes the Si/O interface and explodes significantly later. Note that among the models we present here between 12 and 15 M$_{\odot}$ with compactness $\sim$0.6 many do not explode. Models with both lower and higher compactness do explode. The location of the Si/O interface often best explains the explosion properties (\citealt{wang,tsang2022}); models between 12$-$15 M$_{\odot}$ have weak Si/O interfaces further out whose accretion seems insufficient to inaugurate shock revival. We also note that models which explode later, and accrete for longer, do so generally with higher explosion energies.

\begin{figure*}
    \centering
    \includegraphics[width=0.47\textwidth]{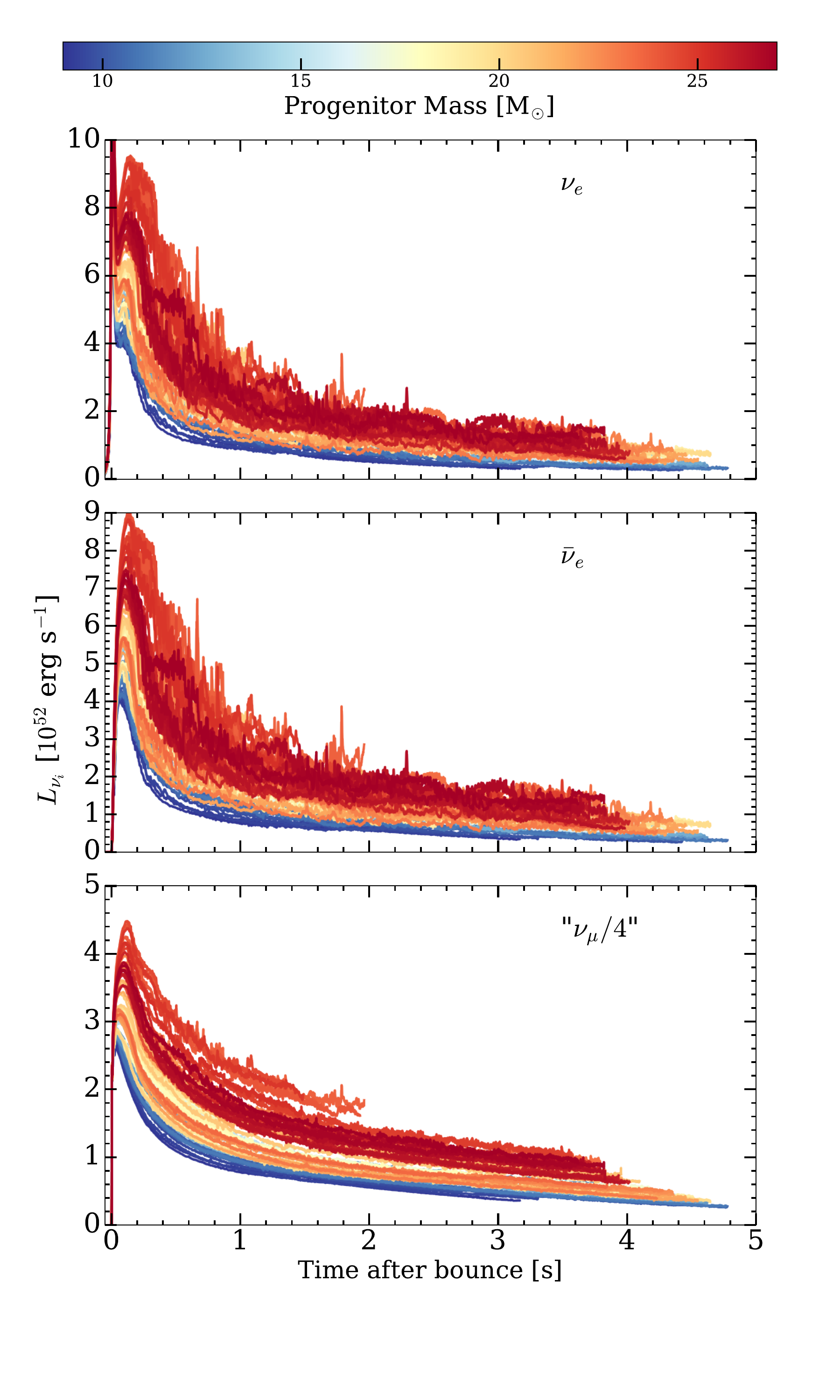}
    \includegraphics[width=0.47\textwidth]{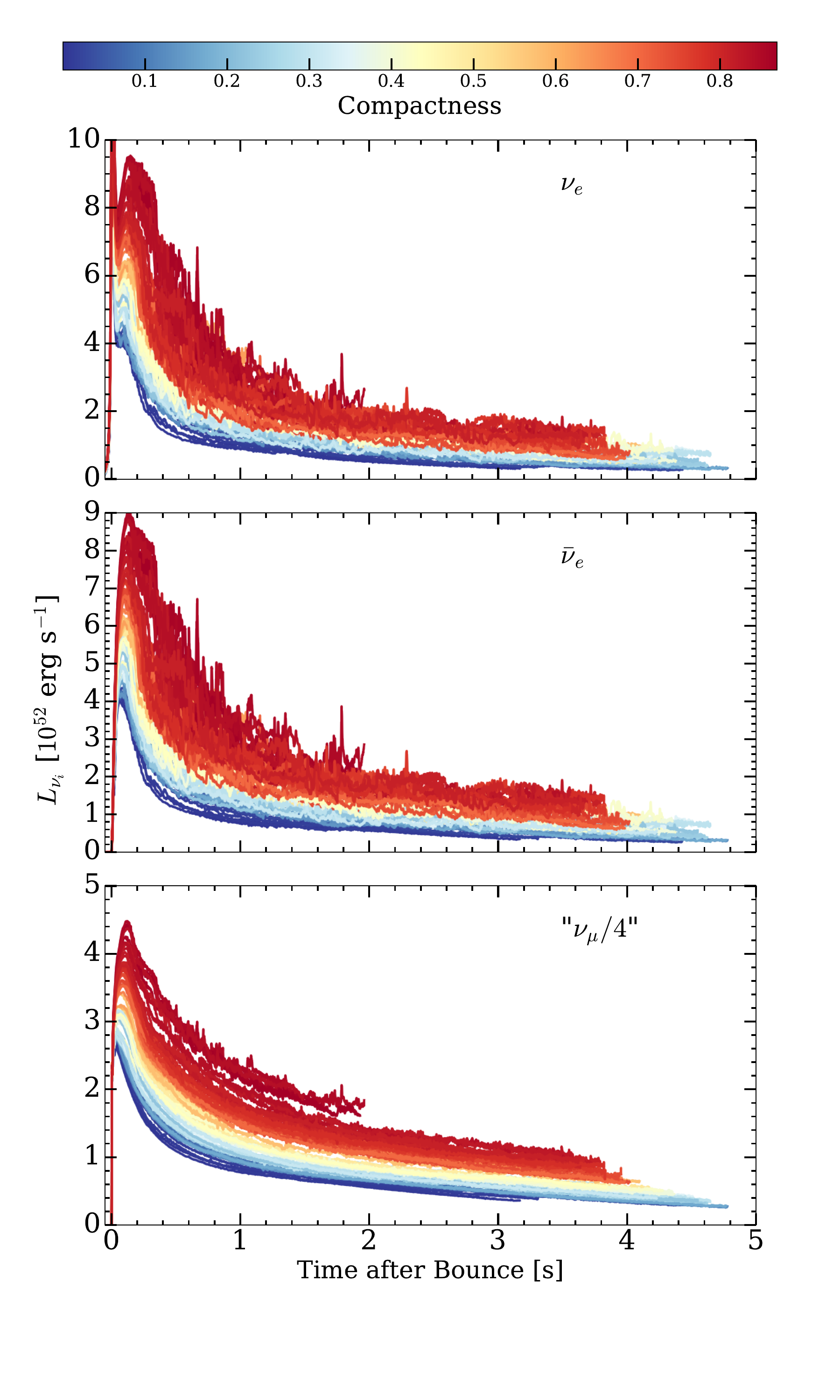}
    \caption{Neutrino luminosity in the observer frame for the three ensemble species at 500 km as a function of time after bounce for our set of 100 2D axisymmetric simulations, colored by progenitor mass (\textbf{left}), and by compactness at 1.75 M$_{\odot}$ (\textbf{right}).}
    \label{fig:lum}
\end{figure*}

In Fig.\,\ref{fig:lum}, we plot the neutrino luminosity at 500 km redshifted (with both velocity and lapse corrections) to the observer frame as a function of time after bounce for all species (the bundled `heavy'-neutrino species' luminosity is divided by 4 to render the luminosity per species). Note that, though we see an approximate correlation with progenitor mass, we see a stronger monotonic correlation with compactness. We see sustained neutrino luminosities of $\sim$10$^{52}$ erg s$^{-1}$ out to $\sim$4-5 seconds after bounce for all species. 
Just several milliseconds prior to the $\nu_e$ breakout burst, there is a smaller $\nu_e$ luminosity peak due to $\nu_e$s produced by the neutronization of the core during collapse \citep{2003ApJ...592..434T,burrows2013,wallace2016}. Typically, the peak electron-neutrino luminosities in the pre-breakout burst are $\sim$10$^{53}$ erg s$^{-1}$. Just before bounce, as the density and temperature of the core increase, the competition between the increase in the opacity and the electron-capture rates leads to a ``mini-peak" (“pre-breakout neutronization peak”) lasting $\sim$milliseconds. The full electron-neutrino breakout which follows is due to electron capture off protons liberated by the nascent shock, lasts $\sim$10 ms, and has typical peak luminosities of $\sim$4.5$\times$10$^{53}$ erg s$^{-1}$. Note that the peak heights and durations for both the breakout and pre-breakout burst depend weakly on the compactness, with higher compactness yielding a slightly longer duration and higher peak.

\begin{figure*}
    \centering
    \includegraphics[width=0.83\textwidth]{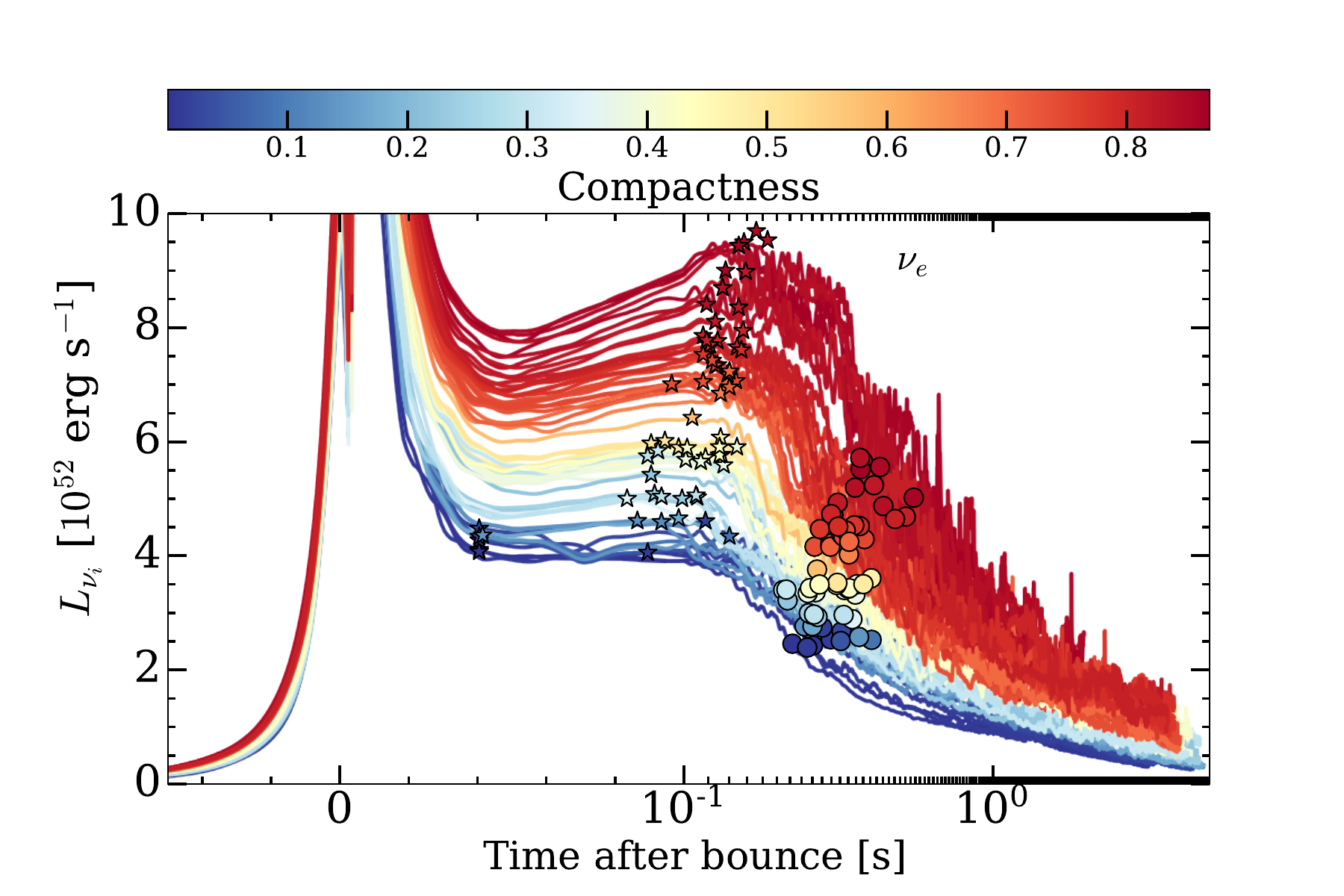}
    \caption{Electron neutrino luminosity for all exploding models. We overplotted as stars where the plateau luminosity is its maximum, and in circles where the luminosity is 60$\%$ of peak. Note the strong correlation between peak luminosity, plateau duration, and compactness.}
    \label{fig:plateau}
\end{figure*}

\begin{figure*}
    \centering
    \includegraphics[width=0.83\textwidth]{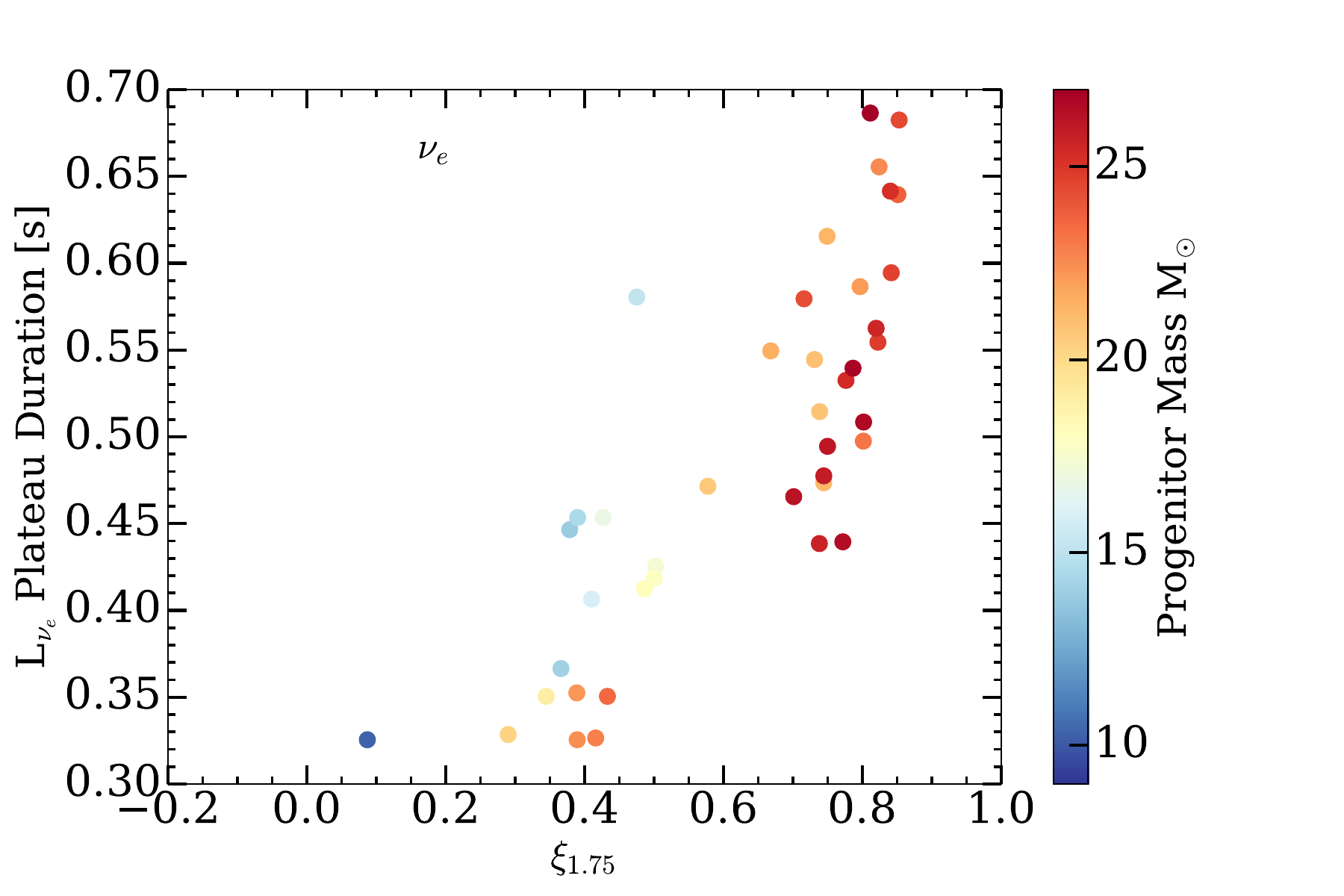}
    \caption{The plateau duration (in seconds) for the electron neutrino luminosity for all exploding models as a function of compactness and colored by progenitor mass. The plateau duration is defined from the end of breakout until when the electron neutrino luminosity drops below 3$\times$10$^{52}$ erg s$^{-1}$. Note the strong correlation between compactness and plateau duration; more compact models sustain longer accretion and higher neutrino luminosities.}
    \label{fig:plateau_corr}
\end{figure*}

In Fig.\,\ref{fig:plateau}, we plot the electron neutrino luminosity during the plateau phase (out to $\sim$several tenths of a second) as a function of time after bounce for all exploding models, and overplot both the peak luminosity and the time at which the electron neutrino luminosity drops to 60$\%$ of the peak value. We see a  strong correlation between peak luminosity and plateau duration, and for both with compactness as well, suggesting a `Phillips-like' relation for CCSN neutrino light curves. Non-exploding models would continue to accrete for longer times. Note that the duration of the plateau also corresponds strongly with the explosion time $-$ explosion reverses accretion, resulting in a noticeable drop in luminosity, often by a factor of several. We emphasize this in Fig.\,\ref{fig:plateau_corr}, where we show the electron neutrino luminosity plateau duration (defined from neutrino breakout until the luminosity drops below 3$\times$10$^{52}$ erg s$^{-1}$) for all exploding models, and highlight the strong correlation with plateau duration and progenitor compactness. In multi-D, we generically expect sustained accretion along certain directions and explosion/ejection along others \citep{vartanyan2018a,2021Natur.589...29B}.

\begin{figure*}
    \centering
    \includegraphics[width=0.47\textwidth]{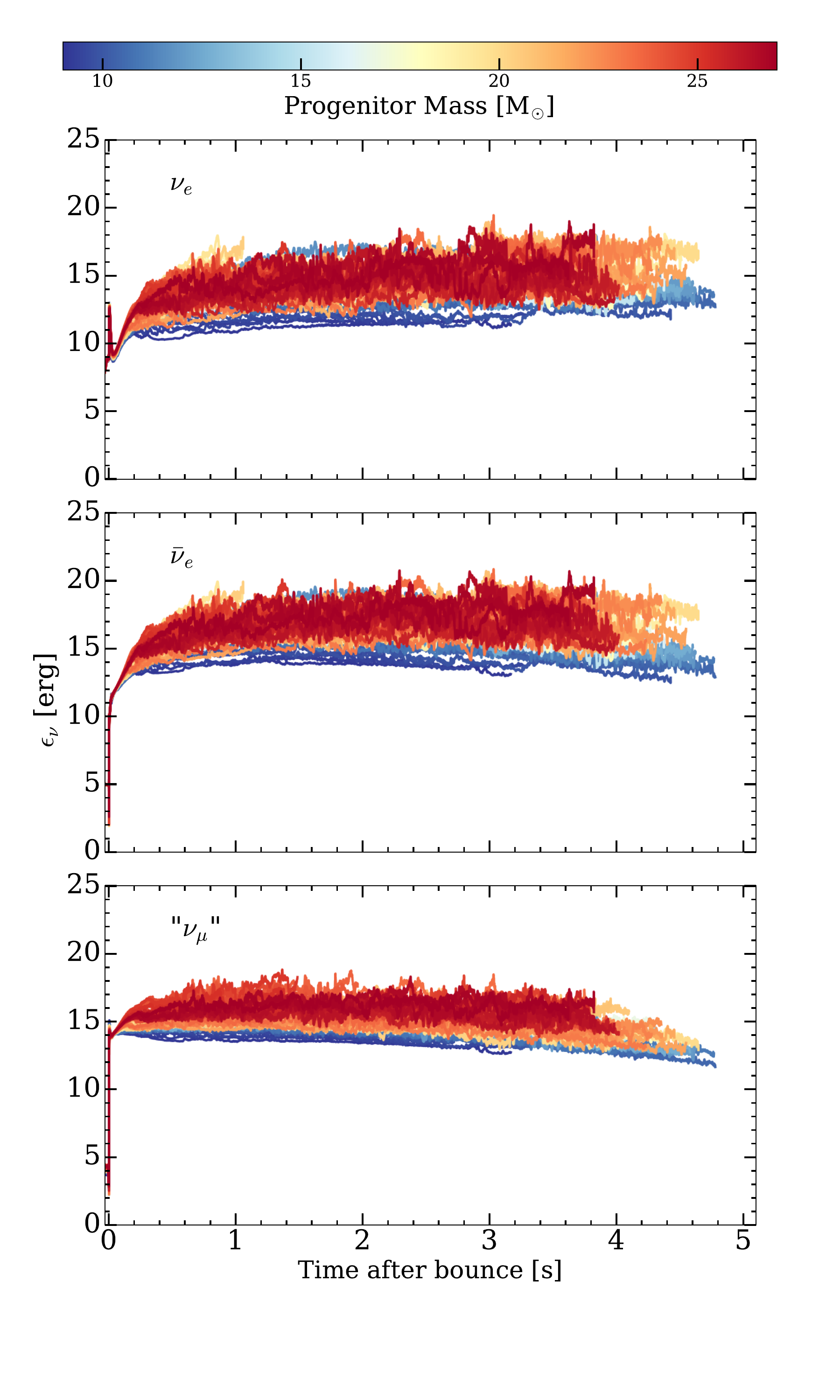}
    \includegraphics[width=0.47\textwidth]{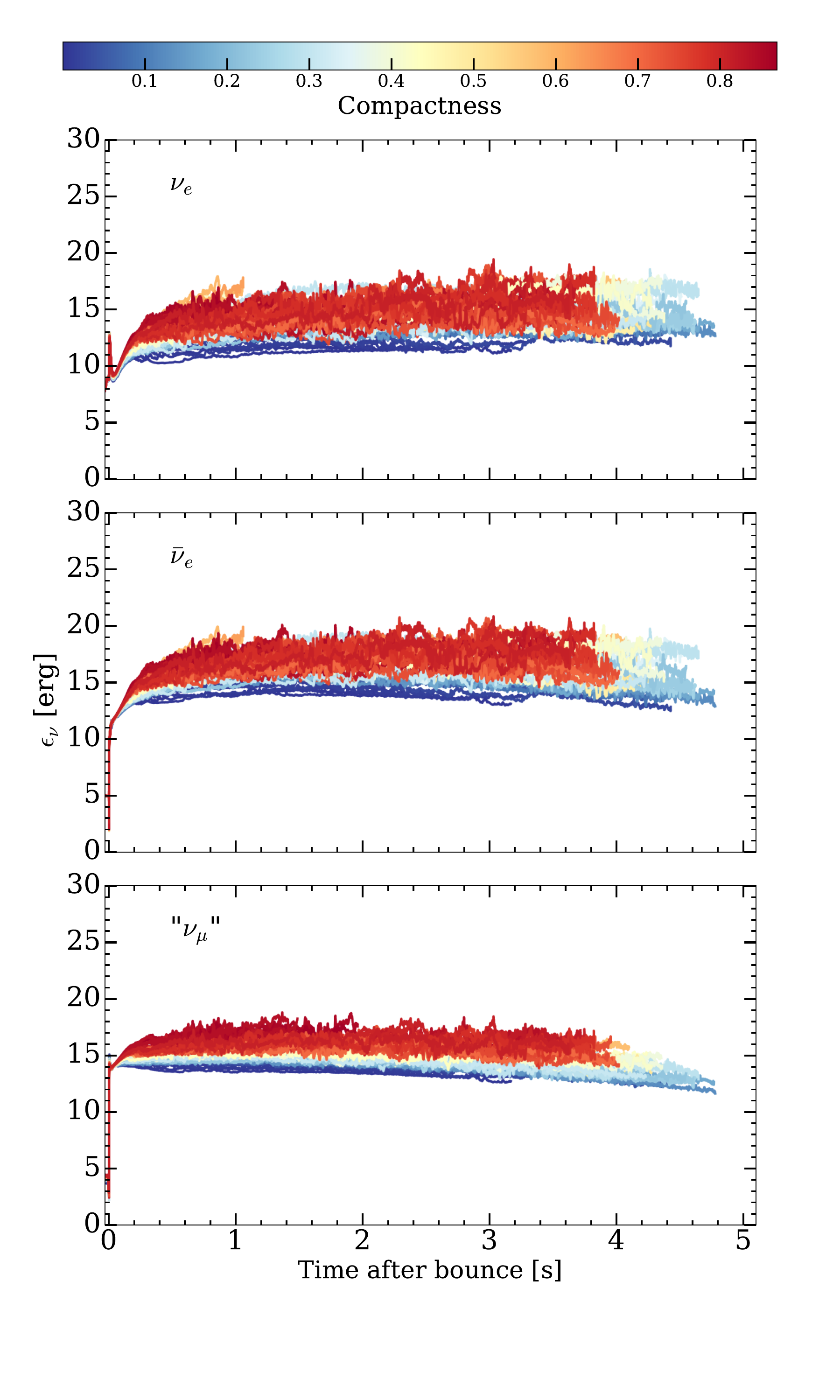}
    \caption{Average neutrino energy in the observer frame for the three ensemble species at 500 km as a function of time after bounce (in seconds) as a function of progenitor mass (\textbf{left}) and compactness (\textbf{right}).}
    \label{fig:eave}
\end{figure*}
\begin{figure*}
    \centering
    \includegraphics[width=0.47\textwidth]{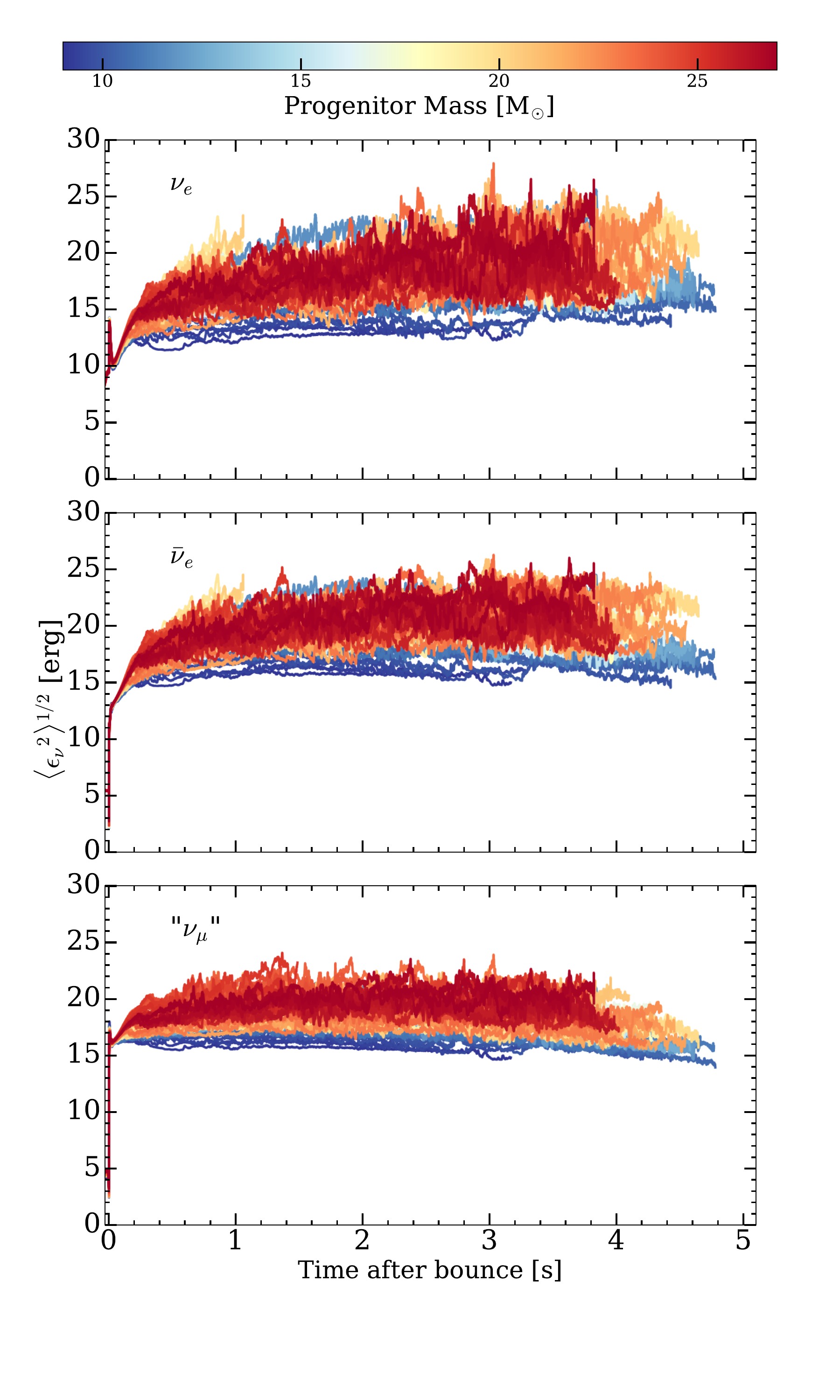}
    \includegraphics[width=0.47\textwidth]{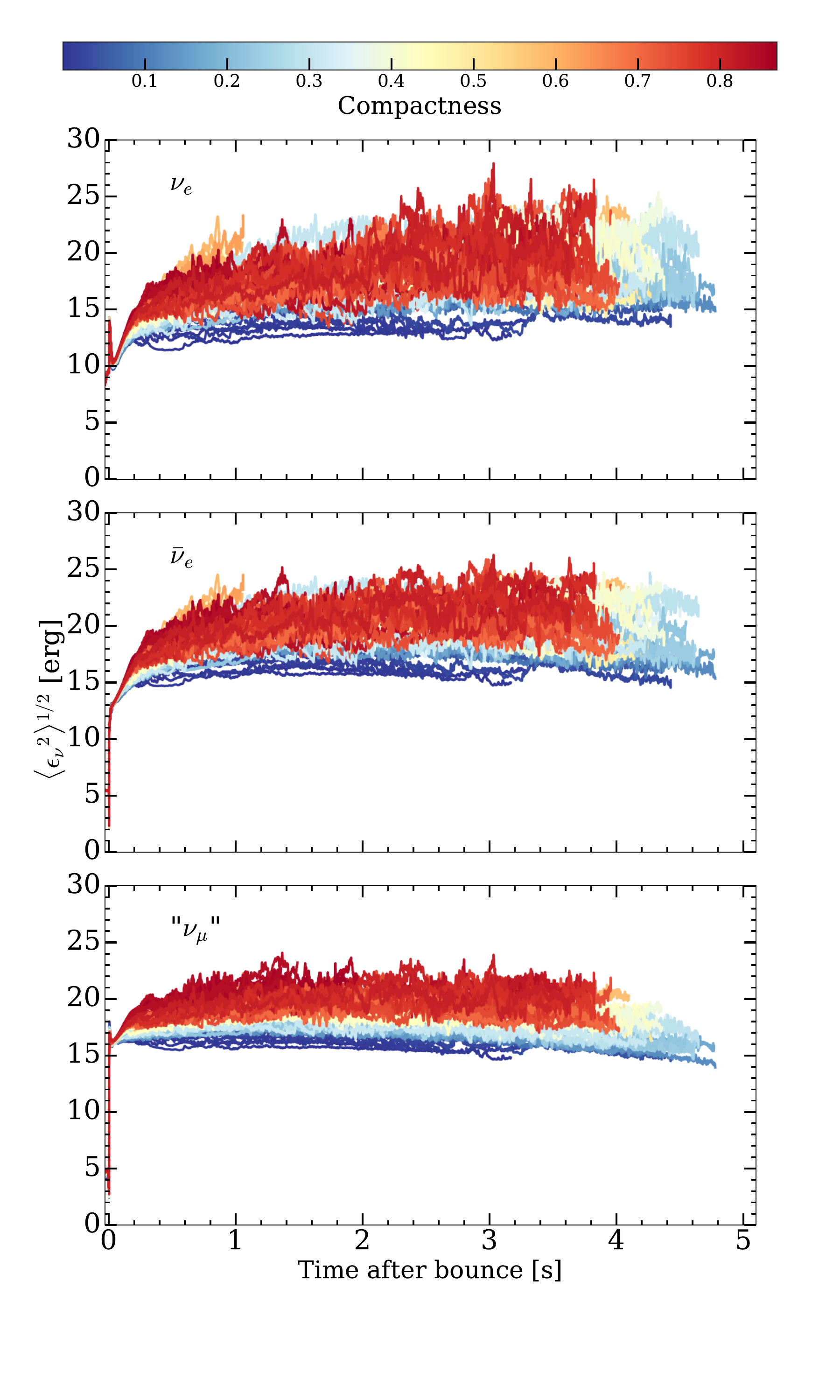}
    \caption{Same as Fig.\,\ref{fig:eave}, but for the RMS neutrino energy.}
    \label{fig:erms}
\end{figure*}
\begin{figure*}
    \centering
    \includegraphics[width=0.32\textwidth]{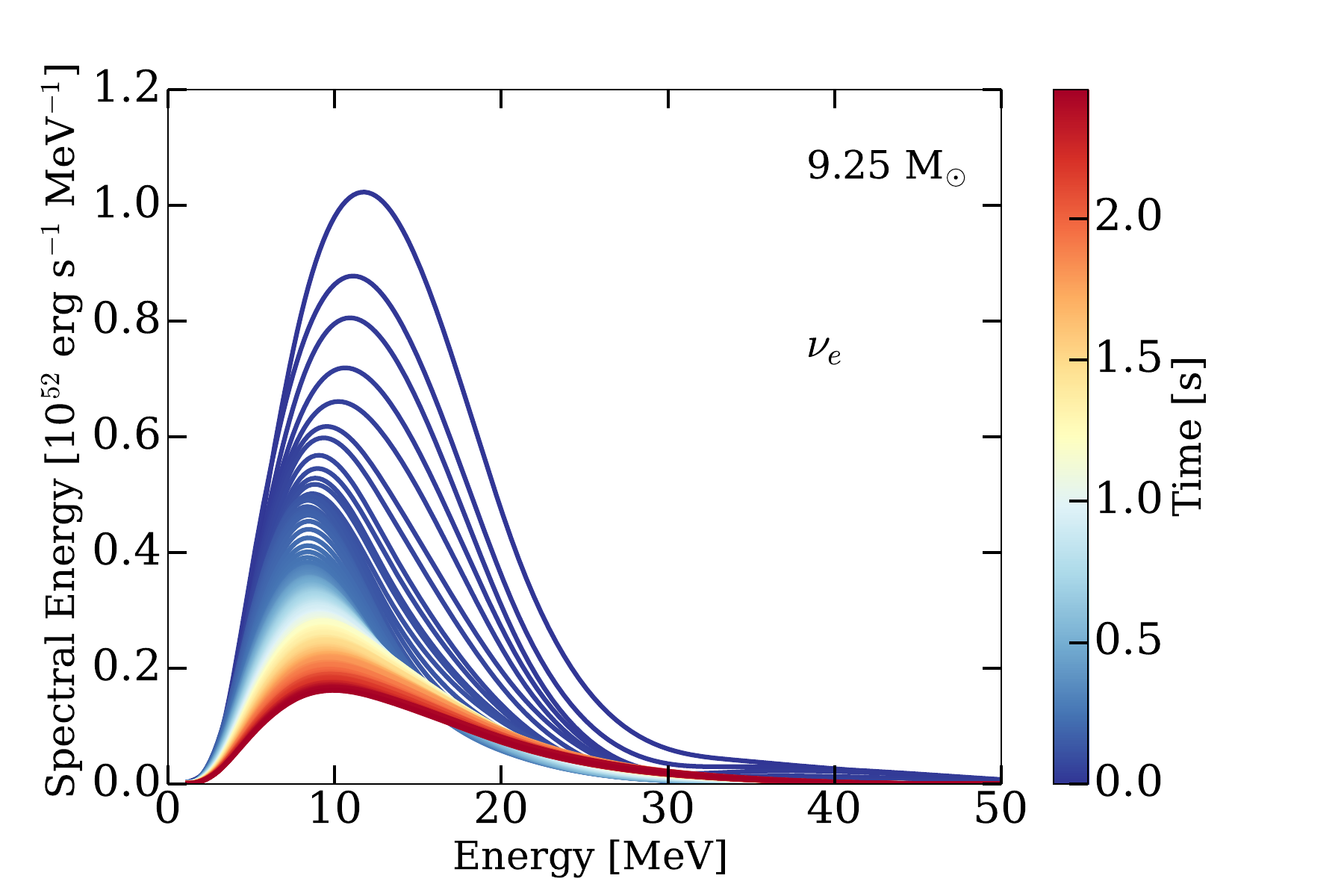}
    \includegraphics[width=0.32\textwidth]{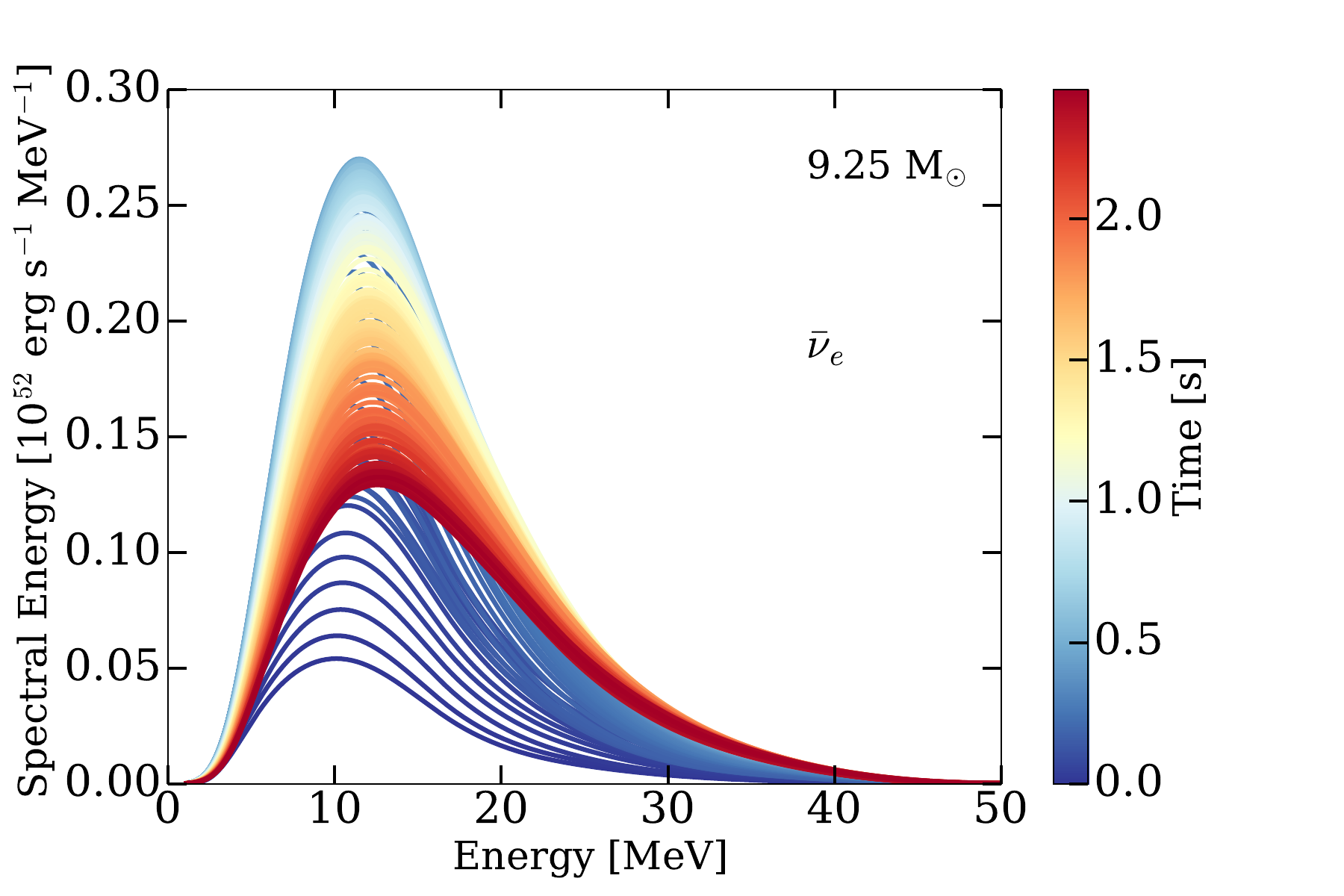}
    \includegraphics[width=0.32\textwidth]{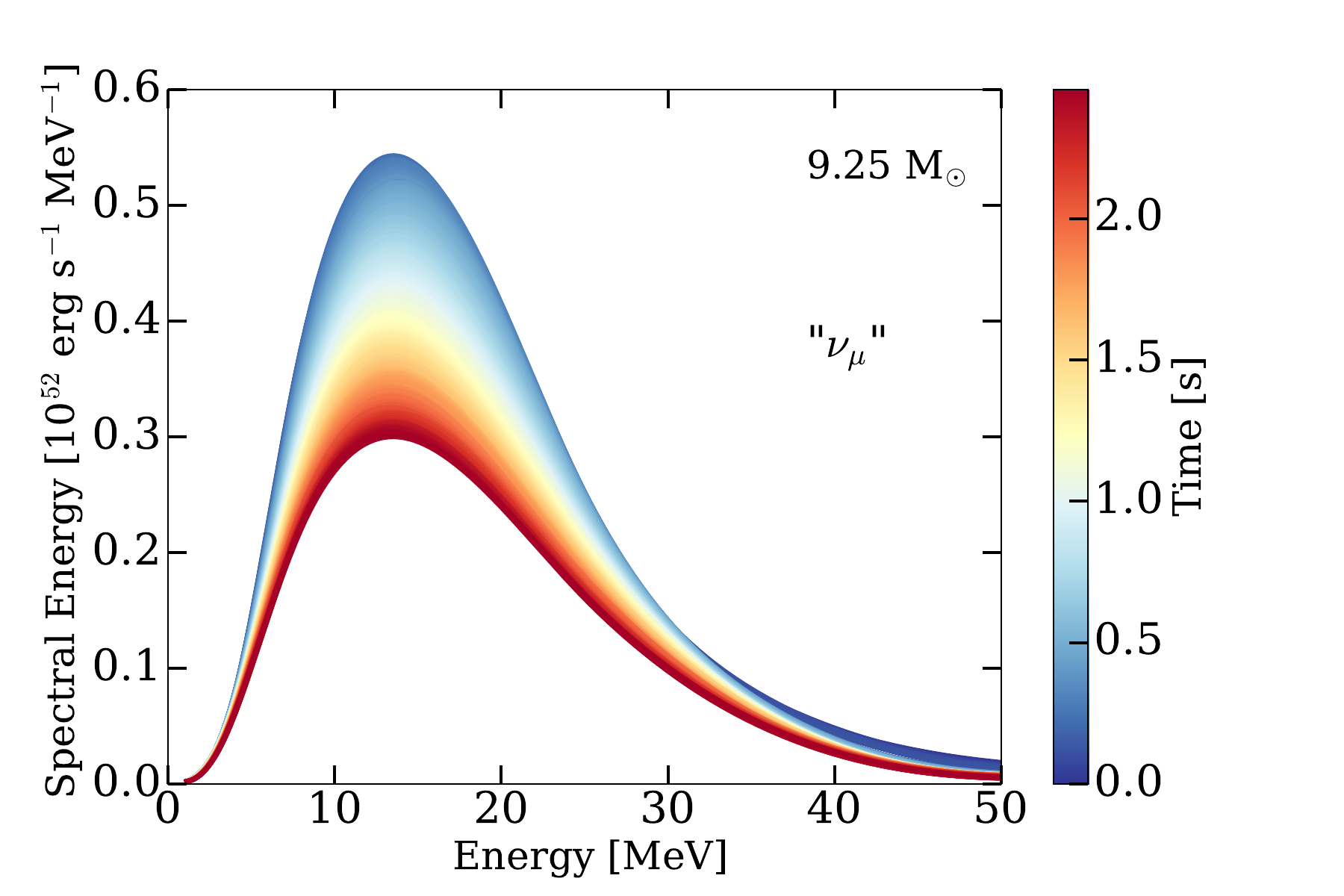}
    \caption{Neutrino spectra for all species as a function of time after bounce for the 9.25 M$_{\odot}$ progenitor.  We have splined our energy group gridding for these plots.}
    \label{fig:nu925}
\end{figure*}
\begin{figure*}
    \centering
    \includegraphics[width=0.32\textwidth]{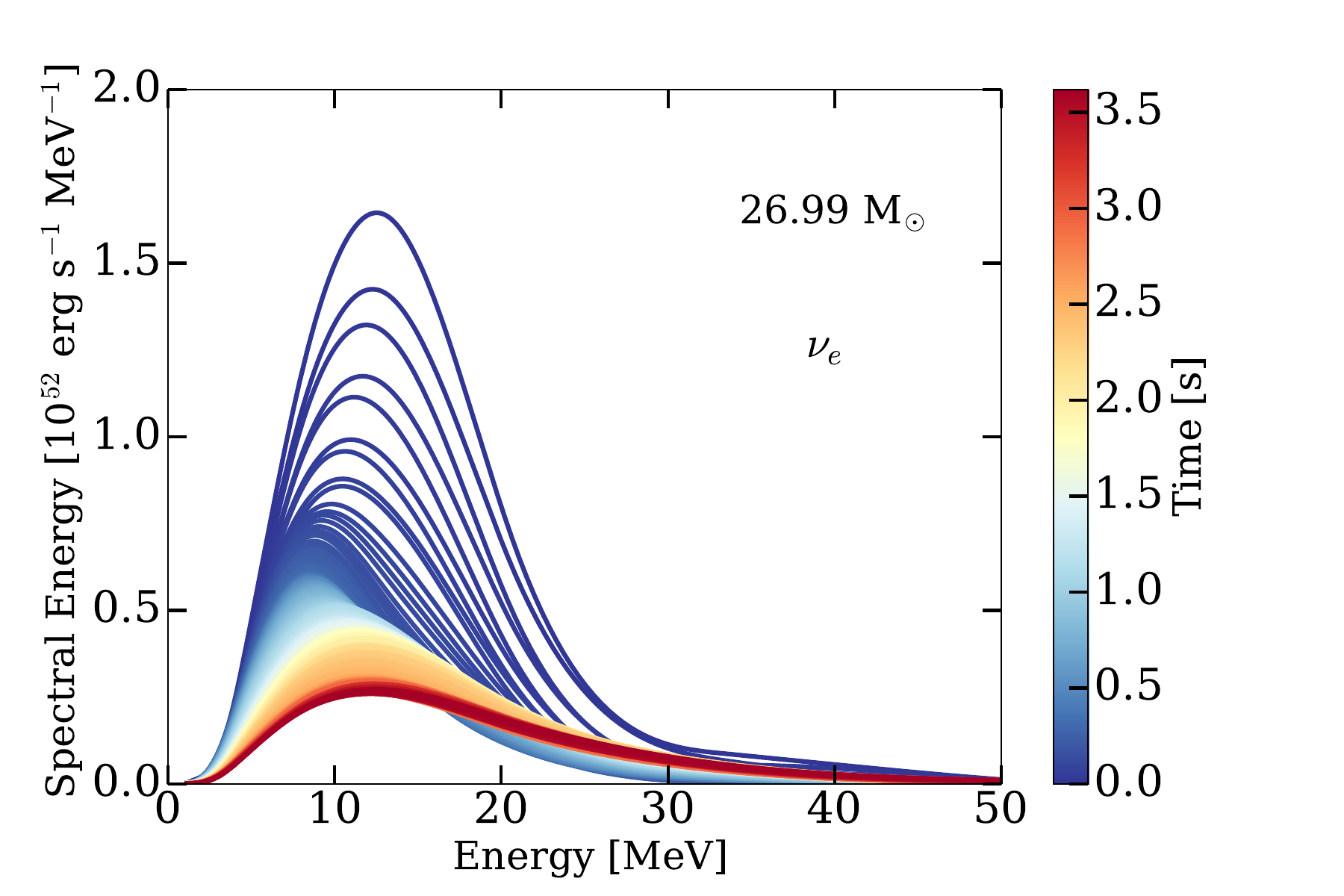}
    \includegraphics[width=0.32\textwidth]{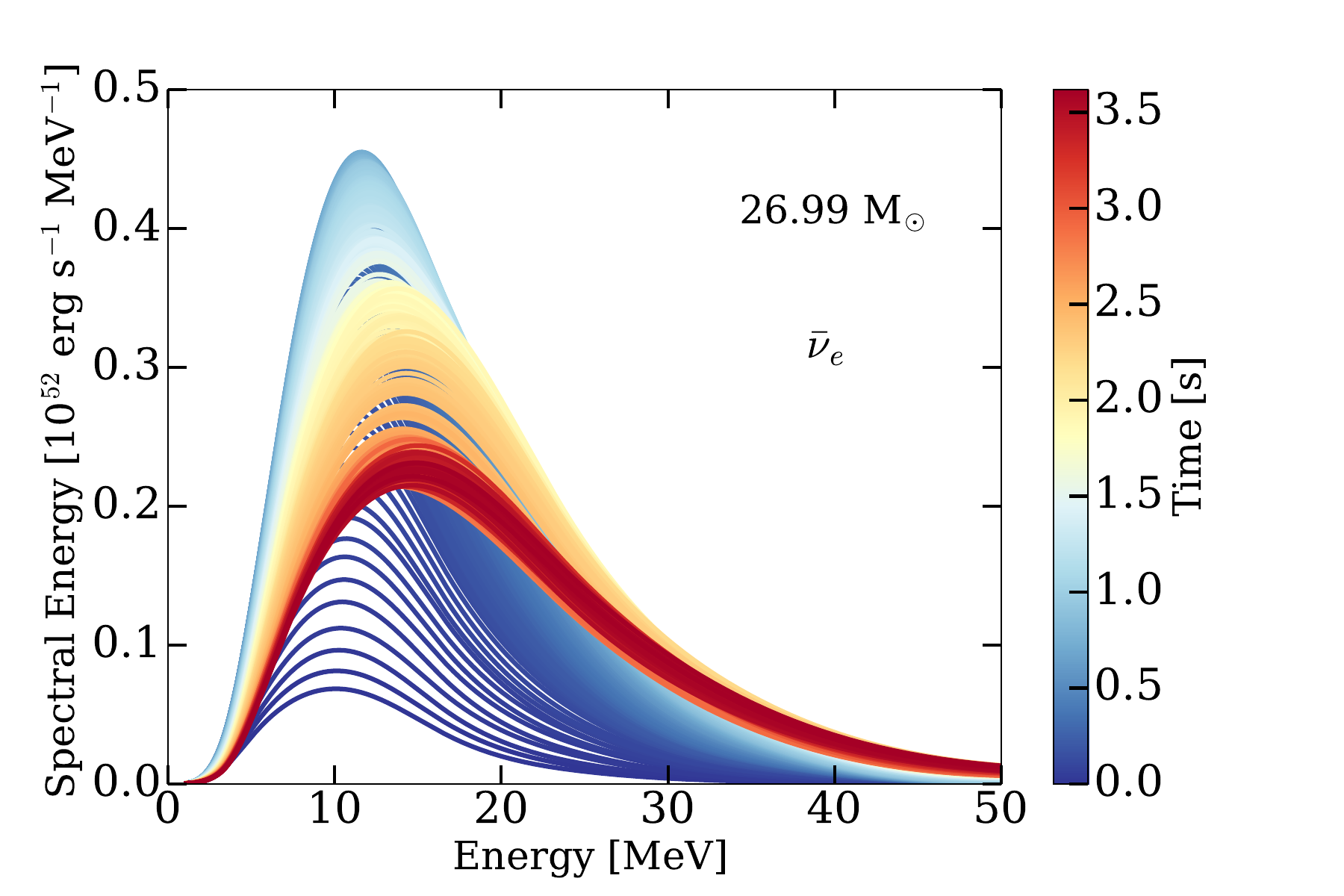}
    \includegraphics[width=0.32\textwidth]{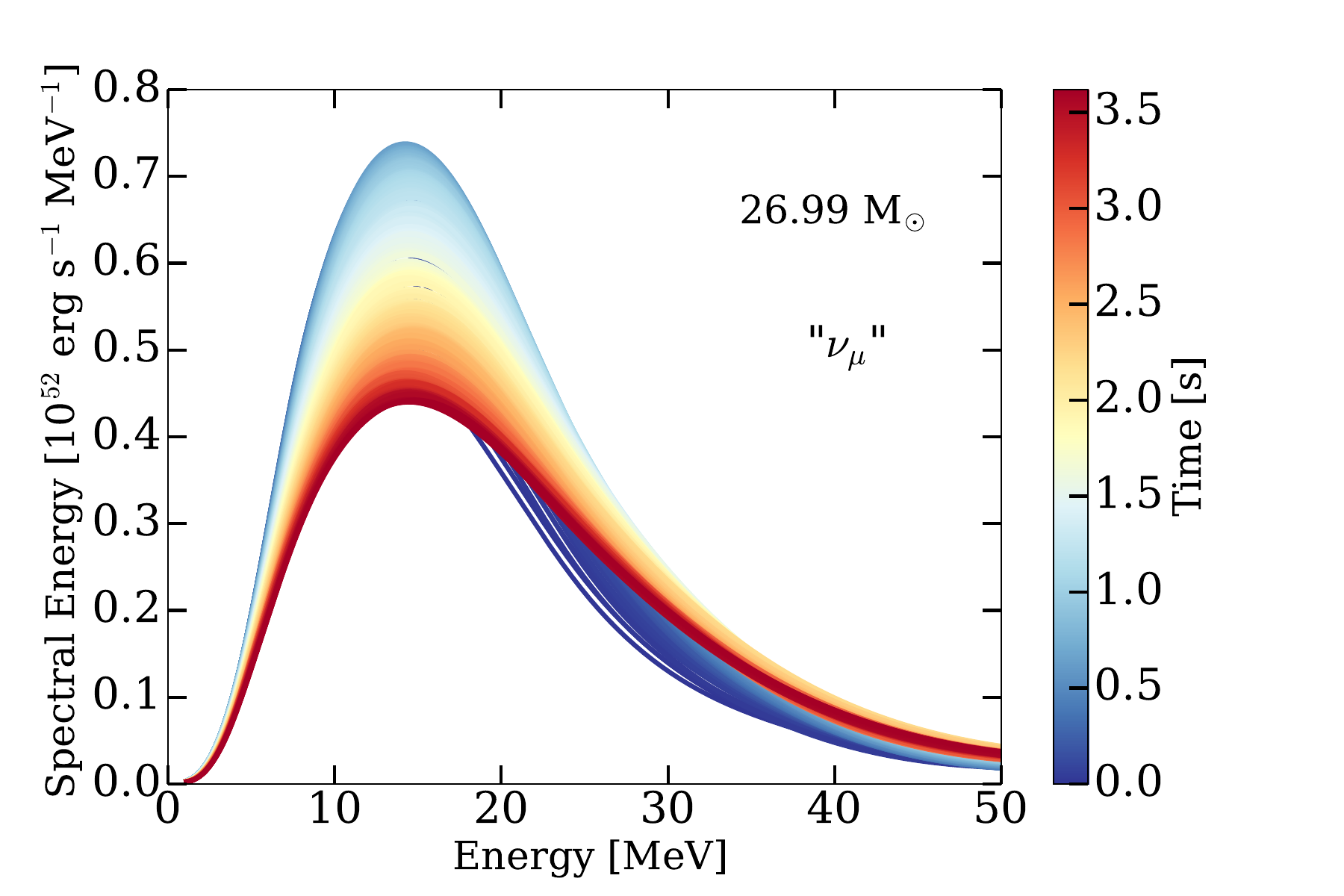}
    \caption{Same as Fig.\,\ref{fig:nu925}, but for the 26.99 M$_{\odot}$ progenitor.}
    \label{fig:nu2699}
\end{figure*}

In Fig.\,\ref{fig:eave}, we show the mean neutrino energies for all species at 500 km (again redshifted to the observer frame). Note that we see a better correlation with compactness than with progenitor mass. In addition, note that models here which do not explode sustain accretion for longer and, hence, have slightly higher mean neutrino energies. In Fig.\,\ref{fig:erms}, we repeat Fig.\,\ref{fig:eave}, but for the RMS neutrino energy. We draw the same inferences. In Figs.\,\ref{fig:nu925} and \ref{fig:nu2699}, we provide sample spectra for the 9.25- and 26.99-M$_{\odot}$ models for all three species.


\subsection{Temporal Fluctuations}

\begin{figure*}
    \centering
    \includegraphics[width=0.47\textwidth]{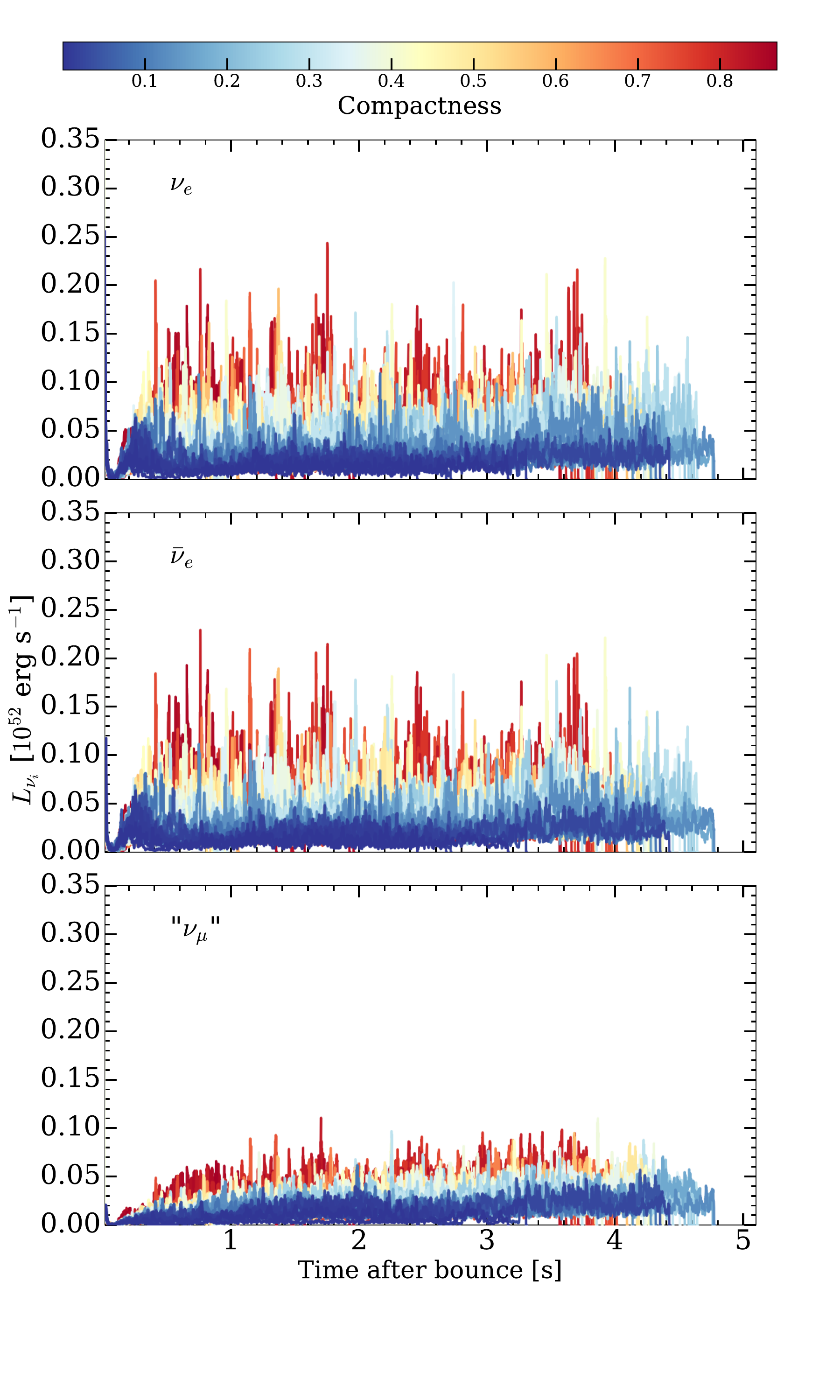}
    \includegraphics[width=0.47\textwidth]{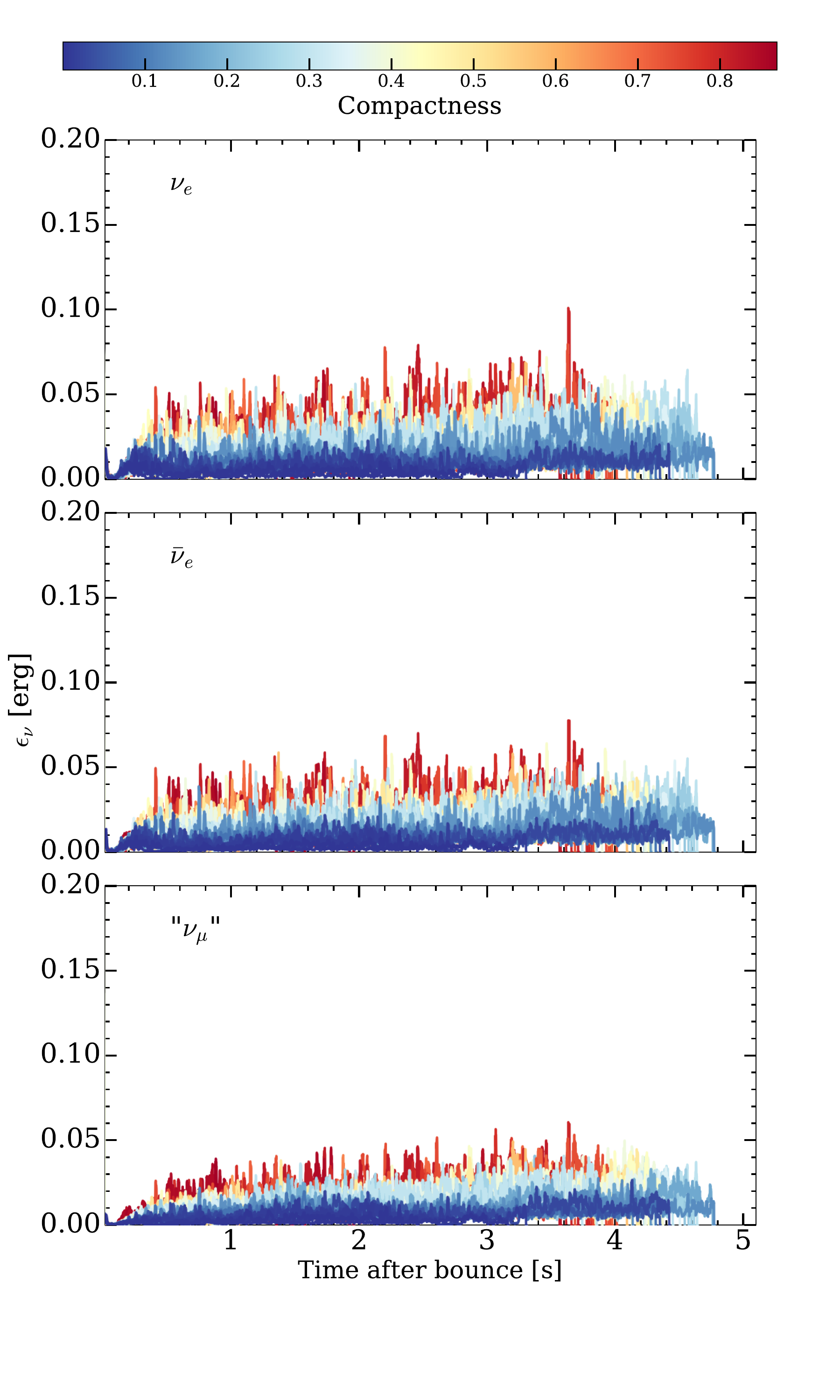}
    \caption{Fractional RMS fluctuation of the neutrino luminosity and the average neutrino energy around a boxcar mean of 5 ms . Note the we see a hierarchy with both compactness (more compact models evince greater fluctuation) and species (electron neutrinos have the most variation, followed by electron anti-neutrinos, and then the bundled `heavy' neutrinos). We also see greater fluctuation in the neutrino luminosity than in the mean neutrino energy.}
    \label{fig:var}
\end{figure*}

With Fig.\,\ref{fig:var}, we briefly explore the fluctuations of neutrino luminosities and mean energies with time. Depicted is the fractional root-mean square (RMS) variation of the neutrino luminosity around a boxcar running time-average of 5 ms. We clearly see a hierarchy with both compactness (more compact models vary more). Furthermore, we see similar variation in time for electron neutrino and electron anti-neutrino quantities, followed by less variation for the bundled `heavy' neutrinos. Note that we found a similar hierarchy earlier (\citealt{vartanyan2019}, see also \citealt{2020ApJ...896..102K}), looking also at the spatial variation of the neutrino diagnostics by species. We see less variation in the neutrino luminosity than the mean neutrino energy. 


\subsection{PNS Mass and Integrated Neutrino Energy Loss}\label{mass_total}
\begin{figure*}
    \centering
    \includegraphics[width=0.87\textwidth]{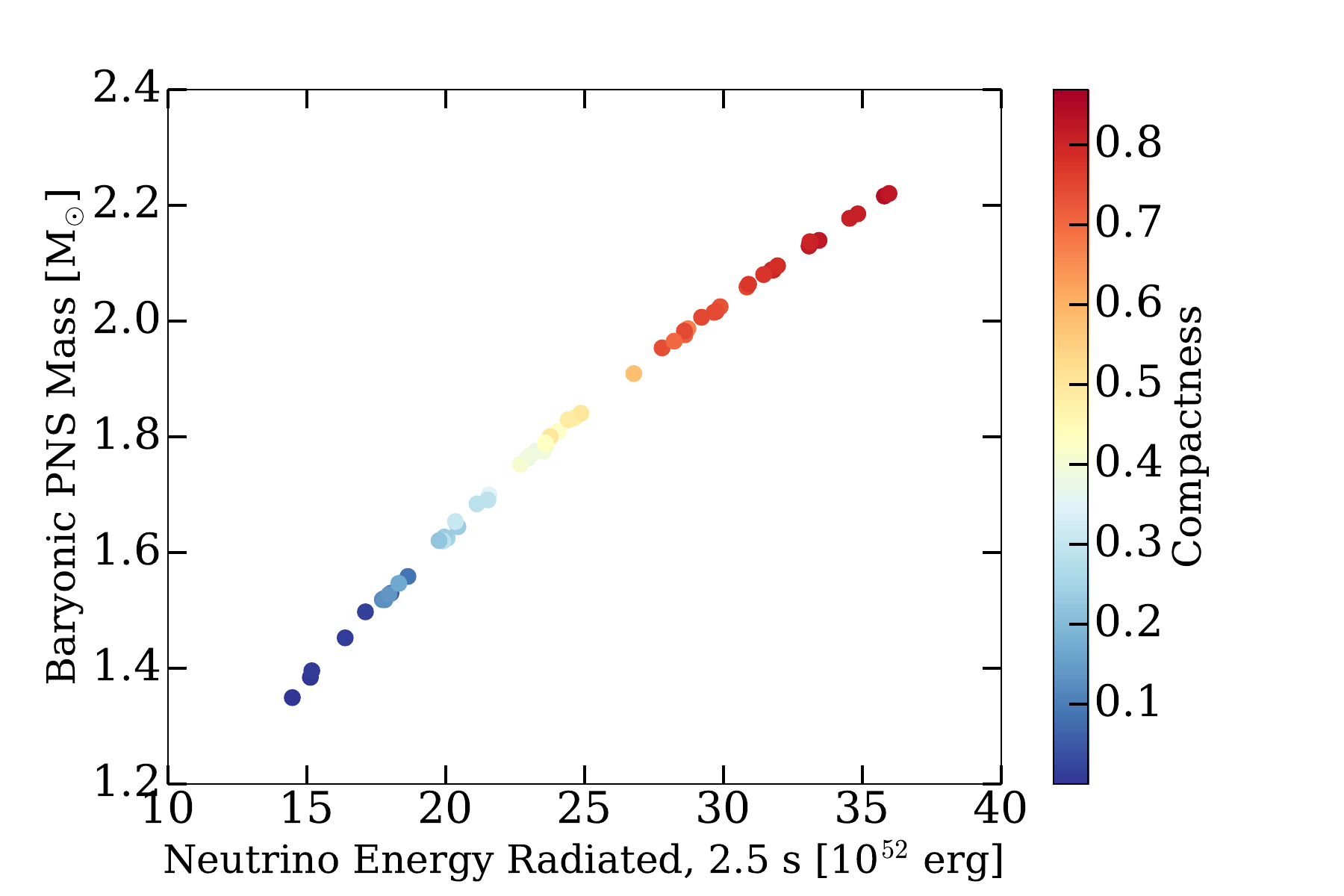}
    \caption{Baryonic PNS masses (M$_{\odot}$) versus the total emitted neutrino energy ( 10$^{52}$ erg) out to 2.5 s post-bounce for all exploding models. Note that we see a strong correlation between the integrated energy loss, the compactness, and the baryonic PNS mass.}
    \label{fig:pns_nue}
\end{figure*}

In Fig.\,\ref{fig:pns_nue}, we plot the radiated neutrino energy (integrating the total luminosity for all species)  \citep{nagakura2021,nagakura2021a,nagakura2022} out to 2.5 s post-bounce for all exploding models. We note an excellent correlation between compactness, baryonic PNS mass, and the integrated energy radiated. More compact models sustain longer accretion that yield higher PNS masses and radiate greater neutrino energy. Observing the radiated neutrino energy  will provide a strong constraint on the PNS mass for CCSN detections (see the quadratic fitting in \citealt{nagakura2022}).




\section{Conclusions}\label{sec:conc}

Traditionally, the supernova neutrino detection community has relied over the decades for the calibration and benchmarking of their subsurface experiments on a heterogeneous collection of supernova explosion models (e.g., \citealt{theia} and \citealt{2022PhRvD.106d3026E}, and references within their Table 1). The latter have emerged haphazardly and rarely comprise a uniform set that might enable the study of physical parameters of the (theoretical) explosion. In addition, no systematic studies have emerged of the neutrino discriminants of microphysical parameters of the neutrino-matter interaction and/or the nuclear equation of state that do not explode models artificially. The construction using the sophisticated numerical tools now available of such a comprehensive forward model set would indeed be a daunting enterprise. There have been large model sets available (e.g., \citealt{2012ApJ...757...69U,2018MNRAS.475.1363H,suwa2019,2021ApJ...909..169K}) employing ad hoc methods to explode 1D models that would not otherwise explode under the current multi-D neutrino-powered/turbulence paradigm of supernova explosions \citep{burrows:95,janka2012,burrows_2020,2021Natur.589...29B}. These studies nevertheless span a useful range of truly physical parameters (such as progenitor mass). However, again these models were not the product of a self-consistent multi-D treatment. With this paper we have attempted to bridge the gap between uniform model sets artificially exploded and self-consistent multi-D (though 2D) datasets that span a wide range of at least one physical quantity (in this case compactness and/or progenitor mass). 

Such a uniform suite of neutrino signals from state-of-the-art CCSN simulations will be, we believe, an essential resource to ensure that the international experiment and modeling communities can most efficiently optimize the retrieval of physical information about the next galactic core-collapse supernova. We must emphasize that our models, though sophisticated, should not be viewed as the final word. Far from it. They 1) do not incorporate the possible effects of neutrino oscillations in the core itself; 2) do not address initial core rotation of significance; 3) rely on a extensive initial progenitor model suite that is a work in progress
for the stellar evolution community; and 4) importantly are performed in axisymmetry. Nevertheless, these one hundred multi-group radiation/hydrodynamic models systematically span an important range of stellar masses, were conducted to significantly later times than is traditional, and were evolved using the same code and microphysics throughout. The result is hoped to be a large collection of models of uniform provenance that will facilitate into the near future the better understanding of core-collapse theory
and modeling among interested experimentalists and 
the further integration of the broader supernova neutrino community.

\clearpage
\newpage
\begin{multicols}{2}

\begin{table*}
\center\caption{\Large{Explosion Properties}}
\begin{tabular}{ |p{1.5cm}||p{1cm}|p{1cm}|p{1cm}|  }
    \hline\hline
  Model        & Run Time &  $\xi_{1.75}$ & Explosion Time  \\
(M$_{\odot}$) & (s, pb) &  & (s, pb)  \\\hline
9.0 & 3.22 & 6.8e-5 & 0.084 \\
9.25 & 2.51 & 0.003 & 0.078 \\
9.5 & 2.73 & 0.009 & 0.07 \\
9.75 & 3.36 & 0.013 & 0.072 \\
10.0 & 2.77 & 0.017 & 0.124 \\
10.25 & 4.48 & 0.026 & 0.08 \\
10.5 & 4.4 & 0.041 & 0.064 \\
10.75 & 4.37 & 0.087 & 0.068 \\
11.0 & 4.83 & 0.125 & 0.1 \\
11.25 & 4.19 & 0.131 & 0.096 \\
11.5 & 4.33 & 0.137 & 0.064 \\
11.75 & 4.82 & 0.17 & 0.122 \\
12.00-BH & 0.94 & 0.301 & - \\
12.03-BH & 0.73 & 0.395 & - \\
12.07-BH & 1.02 & 0.332 & - \\
12.1-BH & 3.91 & 0.297 & - \\
12.13 & 4.67 & 0.249 & 0.114 \\
12.15 & 4.65 & 0.237 & 0.128 \\
12.18-BH & 0.99 & 0.342 & - \\
12.20-BH & 0.92 & 0.322 & - \\
12.25 & 4.63 & 0.236 & 0.134 \\
12.33-BH & 0.96 & 0.344 & - \\
12.40-BH & 0.94 & 0.381 & - \\
12.45-BH & 0.94 & 0.376 & - \\
12.50-BH & 0.96 & 0.347 & - \\
12.54-BH & 0.92 & 0.371 & - \\
12.60-BH & 0.91 & 0.439 & - \\
12.63 & 4.29 & 0.378 & 0.174 \\
12.70 & 4.35 & 0.366 & 0.208 \\
12.72-BH & 0.93 & 0.372 & - \\
12.75 & 3.95 & 0.39 & 0.194 \\
12.80-BH & 0.93 & 0.368 & - \\
12.85-BH & 0.95 & 0.464 & - \\
12.90-BH & 0.93 & 0.362 & - \\
12.93 & 4.19 & 0.475 & 0.254 \\
12.97-BH & 0.93 & 0.411 & - \\
13.00-BH & 0.95 & 0.485 & - \\
13.05-BH & 0.92 & 0.51 & - \\
13.11 & 4.42 & 0.41 & 0.156 \\
13.25-BH & 0.94 & 0.527 & - \\
13.27-BH & 0.92 & 0.453 & - \\
13.32-BH & 0.95 & 0.542 & - \\
13.40-BH & 0.95 & 0.532 & - \\
13.45 & 4.31 & 0.427 & 0.156 \\
13.50-BH & 0.98 & 0.375 & - \\
13.60-BH & 0.94 & 0.525 & - \\
13.75 & 4.22 & 0.502 & 0.138 \\
13.82-BH & 0.91 & 0.529 & - \\
13.90-BH & 0.93 & 0.422 & - \\
13.96 & 4.32 & 0.5 & 0.174 \\
\hline
\end{tabular}
\begin{tabular}{ |p{1.5cm}||p{1cm}|p{1cm}|p{1cm}|  }
\hline\hline
  Model        & Run Time  &  $\xi_{1.75}$ & Explosion Time \\
(M$_{\odot}$) & (s, pb)  &  & (s, pb) \\\hline
14.01 & 4.24 & 0.486 & 0.174 \\
14.13-BH & 1.0 & 0.436 & - \\
14.25-BH & 0.98 & 0.418 & - \\
14.40-BH & 0.89 & 0.552 & - \\
14.41-BH & 0.92 & 0.493 & - \\
14.43 & 4.56 & 0.345 & 0.094 \\
14.44-BH & 0.86 & 0.529 & - \\
14.70-BH & 0.92 & 0.566 & - \\
14.87-BH & 0.92 & 0.59 & - \\
15.00-BH & 1.1 & 0.592 & - \\
15.01 & 4.69 & 0.289 & 0.118 \\
15.04-BH & 0.92 & 0.617 & - \\
15.05 & 3.8 & 0.29 & 0.118 \\
15.38-BH & 1.11 & 0.626 & - \\
16.43 & 4.15 & 0.577 & 0.168 \\
16.65 & 4.0 & 0.738 & 0.202 \\
16.99 & 3.74 & 0.731 & 0.208 \\
17.00 & 3.88 & 0.744 & 0.204 \\
17.07 & 4.01 & 0.749 & 0.332 \\
17.10 & 4.01 & 0.668 & 0.212 \\
17.40 & 4.5 & 0.285 & 0.156 \\
17.48 & 4.6 & 0.222 & 0.16 \\
17.50 & 3.87 & 0.797 & 0.272 \\
17.51 & 4.41 & 0.389 & 0.148 \\
17.83 & 4.4 & 0.389 & 0.148 \\
18.04 & 3.63 & 0.824 & 0.262 \\
18.05 & 4.32 & 0.416 & 0.174 \\
18.09 & 4.28 & 0.306 & 0.164 \\
18.10 & 3.85 & 0.801 & 0.274 \\
18.50 & 3.84 & 0.809 & 0.566 \\
19.02 & 3.89 & 0.433 & 0.15 \\
19.56 & 2.01 & 0.851 & 0.39 \\
19.83 & 1.63 & 0.859 & 0.152 \\
19.99 & 1.98 & 0.869 & 0.374 \\
20.08 & 4.02 & 0.716 & 0.208 \\
20.09 & 1.41 & 0.853 & 0.428 \\
20.18 & 3.62 & 0.842 & 0.436 \\
20.37 & 3.62 & 0.822 & 0.288 \\
21.00 & 1.54 & 0.841 & 0.402 \\
21.68 & 1.52 & 0.84 & 0.164 \\
22.00 & 3.83 & 0.776 & 0.37 \\
22.30 & 3.76 & 0.82 & 0.192 \\
22.82 & 4.07 & 0.738 & 0.324 \\
23.00 & 3.98 & 0.744 & 0.146 \\
23.04 & 3.84 & 0.75 & 0.352 \\
23.43 & 4.03 & 0.701 & 0.382 \\
24.00 & 3.88 & 0.772 & 0.248 \\
25.00 & 3.71 & 0.802 & 0.206 \\
26.00 & 3.87 & 0.786 & 0.242 \\
26.99 & 3.67 & 0.811 & 0.57 \\
\hline
  \end{tabular}

  \begin{flushleft}\large{Table of our 2D simulation results. We include the model name (indicating black hole formers with -BH), the run time in seconds post-bounce, the compactness $\xi_{1.75}$ at 1.75 M$_{\odot}$, and the explosion time in seconds post-bounce.} \end{flushleft}
  \label{sn_tab}
\end{table*}

\end{multicols}

\clearpage
\newpage

\section*{Data Availability}

The data underlying this article are publicly available at \url{https://dvartany.github.io/data/} and \url{http://www.astro.princeton.edu/~burrows/nu-emissions.2d.large/}.

\section*{Acknowledgements}

We thank Benny Tsang, Tianshu Wang, Hiroki Nagakura, Matthew Coleman, and Christopher White for valuable discussion throughout the course of this project. DV acknowledges support from the NASA Hubble Fellowship Program grant HST-HF2-51520. AB acknowledges support from the U.~S.\ Department of Energy Office of Science and the Office of Advanced Scientific Computing Research via the Scientific Discovery through Advanced Computing (SciDAC4) program and Grant DE-SC0018297 (subaward 00009650) and support from the U.~S.\ National Science Foundation (NSF) under Grants AST-1714267 and PHY-1804048 (the latter via the Max-Planck/Princeton Center (MPPC) for Plasma Physics). The three-dimensional simulations were performed on Blue Waters under the sustained-petascale computing project, which was supported by the National Science Foundation (awards OCI-0725070 and ACI-1238993) and the state of Illinois. Blue Waters was a joint effort of the University of Illinois at Urbana--Champaign and its National Center for Supercomputing Applications. We also acknowledge access to the Frontera cluster (under awards AST20020 and AST21003), and this research is part of the Frontera computing project at the Texas Advanced Computing Center \citep{Stanzione2020}. Frontera is made possible by NSF award OAC-1818253. Additionally, a generous award of computer time was provided by the INCITE program, enabling this research to use resources of the Argonne Leadership Computing Facility, a DOE Office of Science User Facility supported under Contract DE-AC02-06CH11357. Finally, the authors acknowledge computational resources provided by the high-performance computer center at Princeton University, which is jointly supported by the Princeton Institute for Computational Science and Engineering (PICSciE) and the Princeton University Office of Information Technology, and our continuing allocation at the National Energy Research Scientific Computing Center (NERSC), which is supported by the Office of Science of the U.~S.\ Department of Energy under contract DE-AC03-76SF00098.

\clearpage

\bibliographystyle{mnras}
\bibliography{References}

\bsp	
\label{lastpage}
\end{document}